\documentclass[fleqn,usenatbib]{mnras}

\usepackage[T1]{fontenc}

\DeclareRobustCommand{\VAN}[3]{#2}
\let\VANthebibliography\thebibliography
\def\thebibliography{\DeclareRobustCommand{\VAN}[3]{##3}\VANthebibliography}

\usepackage{graphicx}	%
\usepackage{amsmath}	%
\usepackage{amssymb}	%
\usepackage{newtxtext,newtxmath}

\usepackage{booktabs}
\usepackage{siunitx} %
\DeclareSIUnit[]\solarmass
{\text{\ensuremath{\textup{M}_{\odot}}}}
\DeclareSIUnit[]\solarluminosity
{\text{\ensuremath{\textup{L}_{\odot}}}}
\DeclareSIUnit[]\solarradius
{\text{\ensuremath{\textup{R}_{\odot}}}}
\DeclareSIUnit[]\year
{\text{yr}}
\DeclareSIUnit[]\au
{\text{AU}}
\DeclareSIUnit[]\parsec
{\text{pc}}
\DeclareSIUnit[]\erg
{\text{erg}}
\DeclareSIUnit[]\arcsecond
{\text{as}}

\newcommand{\ts}{\textsuperscript}

\newcommand{\maxdust}{\ensuremath{\dot{\text{M}}_\text{d,max}}}
\newcommand{\avgdust}{\ensuremath{\dot{\text{M}}_\text{d,avg}}}

\newcommand{\swr}{\ensuremath{_{\text{WR}}}}

\newcommand{\rms}[1]{\ensuremath{_{\text{#1}}}}

\title[Dust growth simulations in WCd systems]{An exploration of dust grain growth within WCd systems using an advected scalar dust model}

\author[J. W. Eatson, J. M. Pittard \& S. Van Loo]{
J. W. Eatson,$^{1,2}$\thanks{E-mail: \texttt{\href{mailto:py13je@leeds.ac.uk}{py13je@leeds.ac.uk}}}
J. M. Pittard$^{1}$
and
S. Van Loo$^{1,3}$
\\
$^{1}$School of Physics and Astronomy, University of
       Leeds, Woodhouse Lane, Leeds LS2 9JT, UK\\
$^{2}$Department of Physics and Astronomy, The University of Sheffield, Hicks Building, Hounsfield Road, Sheffield, S3 7RH, UK\\
$^{3}$Department of Applied Physics, Ghent University, Sint-Pietersnieuwstraat 41, Technicum blok 4
9000 Gent, Belgium
}

\date{Accepted XXX. Received YYY; in original form ZZZ}

\pubyear{2022}

\begin{document}
\label{firstpage}
\pagerange{\pageref{firstpage}--\pageref{lastpage}}
\maketitle

\begin{abstract}
\noindent
Dust production is one of the more curious phenomena observed in massive binary systems with interacting winds.
The high temperatures, UV photon flux and violent shocks should destroy any dust grains that condense.
However, in some extreme cases dust production yields of approximately 30\% of the total mass loss rate of the stellar winds have been observed.
In order to better understand this phenomenon a parameter space exploration was performed using a series of numerical models of dust producing carbon phase Wolf-Rayet (WCd) systems.
These models incorporated a passive scalar dust model simulating dust growth, destruction and radiative cooling.
We find that reasonable dust yields were produced by these simulations.
Significant changes in the dust yield were caused by changing the mass loss rates of the stars, with a greater mass loss rate contributing to increased dust yields.
Similarly, a close orbit between the stars also resulted in higher dust yields.
Finally, a high velocity wind shear, which induces Kelvin-Helmholtz (KH) instabilities and wind mixing, drastically increases the dust yields.

\end{abstract}


\section{Introduction}

Binary systems with colliding stellar winds are a fascinating type of system, capable of producing a variety of peculiar phenomena.
The shocks produced from this wind interaction creates some of the most luminous persistent stellar-mass x-ray sources in the night sky \citep{rossloweSpatialDistributionGalactic2015}.
Within the wind collision region the available mechanical energy can exceed $10^4 \, \si{\solarluminosity}$, producing shocks with post-shock temperatures up to $10^8$ \si{\kelvin}.

In particularly energetic colliding wind binary (CWB) systems with a Wolf-Rayet (WR) star and OB-type partner, dust in the form of amorphous carbon grains has been observed to form \citep{allenInfraredPhotometryNorthern1972}.
This is particularly curious, as the high temperatures, strong shocks and UV luminosities of these systems should result in dust grains being rapidly destroyed through sublimation processes.
These dust forming CWB systems have only been observed to occur if a carbon phase WR star (WC star) is partnered with either another WR star or an OB main sequence star (a WR+OB system).
While the exact methods of dust formation and evolution in these systems are poorly understood, dust formation rates have been observed to be extremely high, up to $10^{-6} \, \si{\solarmass\per\year}$.
This is approximately $36\%$ of the total wind by mass in the case of WR104 \citep{lauRevisitingImpactDust2020}.

Within different colliding wind binary systems, dust may form either continuously or periodically.
The first such observed dust forming system was the episodic dust forming system WR140, first reported by \cite{williamsMultifrequencyVariationsWolfrayet1990} who observed a significant and highly variable infrared excess, consistent with emission from dust grains.
The dust production rate was later found to vary by a factor of 40 over the orbital period of  \SI{7.9}{\year} \citep{van1999wolf,thomasOrbitStellarMasses2021}.
Persistent dust forming systems were subsequently discovered, such as WR104 \citep{tuthill_dusty_1999} and WR98a
\citep{monnierPinwheelNebulaWR1999}.
Whilst the exact mechanism for dust formation is not currently known, there is a strong correlation between periodicity and eccentricity, with less eccentric systems forming dust continuously, while highly eccentric systems exhibit episodic dust formation
\citep{crowther_dust_2003}.
Due to this orbital dependency, it is likely that there is an optimal dust forming separation, where dust can form in large quantities. This could be due to factors such as strong post shock cooling, which is highly dependent on the wind speed and orbital separation.
Additionally, dust may be protected from the bulk of the stellar radiation due to the extremely large degree of extinction that may occur in the dense post-shock environment of radiative shocks \citep{cherchneffDustFormationCarbonrich2015}.

Direct observation of dust forming CWBs and in particular the wind collision region (WCR) is exceptionally difficult for a number of reasons:

\begin{itemize}
  \item WR+OB CWB systems are extremely rare. Of the 667 catalogued WR stars at the time of writing, 106 have been confirmed to be in a binary system \citep{rossloweSpatialDistributionGalactic2015,williamsVariableDustEmission2019}.
  \item A WC star is required for dust formation. No nitrogen sub-type Wolf-Rayet (WN) stars have been observed to form dust.
  \item Not all WC+OB systems are dust producing, limiting the sample size further.
  \item 56 dust forming systems with a known spectral type have been observed overall. Despite producing an extremely large quantity of dust in their local region, they are outnumbered by AGB stars by $\sim 3$ orders of magnitude \citep{ishiharaGalacticDistributionsCarbon2011}.
  \item Galactic CWB systems are comparatively distant from earth. For instance, WR 104, a well-studied system, is $\sim \SI{2.5}{\kilo\parsec}$ distant \citep{soulainSPHEREViewWolfRayet2018} and no WCd systems have been detected at a distance of $< 1 \, \si{\kilo\parsec}$ \citep{rossloweSpatialDistributionGalactic2015}. This prevents observations of these systems at a high spatial resolution.
  \item Grain growth from small nucleation grains is predicted to be very rapid in CWB systems \citep{zubkoPhysicalModelDust1998a}. Therefore studying the initial grain evolution would require observations of extremely high angular resolution.
\end{itemize}

For these reasons, numerical simulations are useful for modelling the growth of dust grains within this unresolved region.
In order to better understand what influences dust production in a CWB system, a parameter space exploration of the wind and orbital parameters was performed.
In particular the orbital separation, mass-loss rate and wind velocity were modified for both stars in order to vary the wind momentum ratio, $\eta$, and the cooling parameter, $\chi$.
The wind momentum ratio is defined as:

\begin{equation}
  \label{eq:paper1-eta}
  \eta = \frac{\dot{\text M}_\text{OB} v^\infty_\text{OB}}{\dot{\text M}_\text{WR}v^\infty_\text{WR}} ,
\end{equation}

\noindent
where $\dot{\text{M}}$ is the mass loss rate of a star and $v^\infty$ is the terminal velocity of a star's outflow.
A low value for $\eta$ indicates that the winds are extremely imbalanced, with the WR typically dominating the wind dynamics of the system.
The wind momentum ratio determines for a given orbital separation, $d_\text{sep}$, the distance from each star to the apex of the wind collision.
We define the terms $r_\text{WR}$ and $r_\text{OB}$, representing the distance from the WR and OB stars to the stagnation point of the WCR:

\begin{subequations}
  \begin{align}
    r_\text{WR} & = \frac{1}{1+\eta^{1/2}} d_\text{sep} , \\
    r_\text{OB} & = \frac{\eta^{1/2}}{1+\eta^{1/2}} d_\text{sep} .
  \end{align}
\end{subequations}

\noindent
This assumes the winds both accelerate to terminal speed and that there is no radiative inhibition \citep{stevens_stagnation-point_1994} or braking \citep{gayley_sudden_1997}.
In some systems the winds may be so imbalanced that the stronger wind collides directly with the companion star.
The half-opening angle of the WCR can be estimated by the formula

\begin{equation}
  \theta_c \simeq 2.1 \left( 1 - \frac{\eta^{2/5}}{4}\right) \eta^{-1/3} ~~~ \text{for} ~ 10^{-4} \leq \eta \leq 1 ,
\end{equation}

\noindent
to a relatively high degree of accuracy \citep{eichler_particle_1993,pittardCollidingStellarWinds2018}.

The cooling parameter, $\chi$, compares the cooling time to the escape time from the shocked region for a parcel of gas in the immediate post-shock environment. An approximation can be made using the known parameters of a system using the equation:

\begin{equation}
    \label{eq:paper1-chi}
    \chi = \frac{t_\text{cool}}{t_\text{esc}} \approx \frac{v_8^4 d_{12}}{\dot{\text M}_{-7}} , 
\end{equation}

\noindent
where $v_8$ is the wind terminal velocity in units of $10^8$ \si{cm.s^{-1}}, $d_{12}$ is the distance to the WCR apex in units of $10^{12}$ \si{cm}, and $\dot{\text M}_{-7}$ is the mass loss rate in units of $10^{-7} \si{\solarmass\per\year}$ \citep{stevens_colliding_1992}.
$\chi \leq 1$ indicates that radiative cooling is very important, while $\chi \gg 1$ indicates that the system is adiabatic.
Strong cooling is aided with slow, dense winds and a high metallicity.
As such in many systems the post-shock WR flow will rapidly cool from the immediate post-shock temperature of $\sim 10^{7-8} \, \si{\kelvin}$ to temperatures in the dust formation range, $T \lesssim 10^4 \, \si{\kelvin}$.
A strongly radiating WCR can also be significantly compressed far more as it loses energy.
In comparison, an adiabatic WCR is limited to a maximum density increase of a factor of 4 above the pre-shock wind density for a ratio of specific heats, $\gamma = 5/3$.
The density increase and cool temperatures result in rapid dust growth and protection from the stellar UV radiation in some systems.
Note also that Eq. \ref{eq:paper1-chi} takes account of gas and plasma cooling only, but other cooling, such as dust cooling, may also be important.

In this paper, we aim to explore how dust growth is affected by the orbital and wind parameters of persistent dust forming WR+OB systems.
This is performed by running a series of hydrodynamical simulations with an advected scalar dust model.
In Section \ref{sec:methodology} we outline the methodology of our simulations, and how our dust model is implemented. 
We discuss our model series parameters, and why these parameters were chosen in Section \ref{sec:p1-model-parameters}.
Finally we discuss our results and conclude in Sections \ref{sec:p1-results} and \ref{sec:p1-conclusion}.

\section{Methodology}
\label{sec:methodology}

Numerical simulations within this paper utilise the Athena++ hydrodynamical code, a highly modular fluid dynamics code \citep{stoneAthenaAdaptiveMesh2020}.
Simulations are generated in 3D and the Euler hydrodynamical equations are solved in the form:

\begin{subequations}
  \begin{align}
    \frac{\partial\rho}{\partial t} & + \nabla \cdot \left(\rho \boldsymbol{u}\right) = 0 , \\
    \frac{\partial \rho \boldsymbol{u}}{\partial t} & + \nabla \cdot \left(\rho \boldsymbol{u} u + P \right) = 0, \\
    \frac{\partial \rho \varepsilon}{\partial t} & + \nabla \cdot \left[ \boldsymbol{u} \left( \rho\varepsilon + P \right) \right] = \mathcal{L}\rms{T} , 
  \end{align}
\end{subequations}

\noindent
where $\varepsilon$ is the total specific energy ($\varepsilon = \boldsymbol{u}^2/2 + e/\rho $), $\rho$ is the gas density, $e$ is the internal energy density, $P$ is the gas pressure, $\boldsymbol{u}$ is the gas velocity and $\mathcal{L}\rms{T}$ is the energy loss rate per unit volume from the fluid due to gas and dust cooling.

Athena++ has been configured to run using a piecewise linear reconstruction method with a 4\ts{th} order Strong Stability Preserving Runge-Kutta time-integration method \citep{spiteriNewClassOptimal2002}.
Athena++ was forked from the original repository and additional routines were written for a colliding wind binary scenario.
Routines were created to produce a steady outflow from a small spherical region around a set of cartesian co-ordinates as well as a function to move these co-ordinates with each time-step; these were used to simulate stellar wind outflow and orbital motion, respectively.
Additionally, Athena++ was further modified to include an advected scalar dust model for simulating dust growth and destruction as well as a photon emission cooling model to approximate cooling for gas and dust particles within the fluid.

\begin{figure}
  \centering
  \includegraphics[width=2.5in]{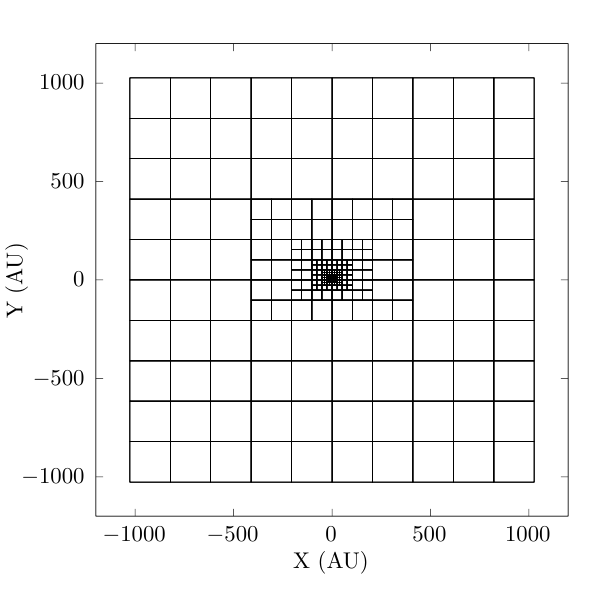}
  \caption[Static mesh refinement example]{A plot of the blocks used in a 7 level simulation with a block size of $32\times 32 \times 8$ cells. The block density increases dramatically closer to the barycentre. The coarse simulation resolution is $(320\times 320\times 40)$ cells with a block size of $(32\times32\times8)$ cells. The diagram is sliced about the $z$ axis at $z=0$.}
  \label{fig:smr-grid}
\end{figure}

Athena++ utilises OpenMPI for parallelism, breaking the simulation into blocks, which are distributed between processors.
The block size is variable, but for these simulations a block size of $32\times 32 \times 8$ was found to be optimal.
This meshblock system is also utilised in mesh refinement for increasing the effective resolution.
As the CWB systems are being simulated in their entirety, a very large volume needs to be simulated, while at the same time the region between the stars must be resolved with a resolution of at least 100 cells in order to adequately resolve the WCR.
This difference in length scales necessitates the use of static mesh refinement (SMR) to improve the effective resolution of the simulation.
A base coarse resolution of $320 \times 320 \times 40$ cells in $XYZ$ is defined for the simulations, while a region close to the stars operates at a higher refinement level.
This rectangular region encompasses the entire orbital path of the stars and is refined to the maximum level.
This results in a resolution increase of a factor of $2^{n-1}$ greater than the coarse resolution, where $n$ is the refinement level (see Fig. \ref{fig:smr-grid}).
In the case of 7 levels (inclusive of the base, ``coarsest'' level) as used in most of the simulations in this paper, this results in an effective resolution of $20480 \times 20480 \times 2560$ cells.
The maximum refinement level is determined based on the orbital separation of the system in order to correctly resolve the WCR from the stars.
In the case of most of the simulations in this paper, we use a $d\rms{sep}$ of \SI{4}{AU}, 7 refinement levels (inclusive of the base, ``coarsest'' level) and a simulation domain of $1000 \times 1000 \times 125 \, \si{AU}$.
For this typical case we calculate a ``finest'' cell volume of \SI{0.05}{AU^3}, which is sufficient to resolve the orbit of the stars using 80 cells, which we determined through experimentation to be sufficient to correctly resolve the WCR.
An open boundary condition is used, stellar wind that flows out of the numerical domain is removed from the simulation.
SMR is utilised instead of Adaptive Mesh Refinement, a more flexible conditional method, as it has proven to be more reliable for our simulations.
As much of the grain evolution occurs a small distance from the WCR stagnation point, much of the simulation volume can be run at a lower resolution without affecting the simulation results.

The wind outflow from each star is simulated by replacing the conserved variables (density, momentum and energy) within a small region around the expected position of the stars; this region is typically on the order of 6 maximally refined cells in radius.
This rewrite corresponds to a change in density, $\rho \rms R$, pressure, $P \rms R$, and total energy, $E \rms R$, imparted by an outflowing wind, such that:

\begin{subequations}
  \begin{align}
    \rho \rms R & = \frac{\dot M}{(4 \pi r^2 v_\infty)} , \\
    P \rms R    & = \rho \rms R k \rms B T \rms w / \mu m \rms H , \\
    E\rms R   &  = \frac{P \rms R}{\gamma - 1} + \frac{1}{2} \rho \rms R v_\infty^2 ,
  \end{align}
\end{subequations}

\noindent
where $v_\infty$ is the wind velocity as it flows radially from the center of the ``remap zone'', $T \rms w$ is the wind temperature and $r$ is the radial distance from the current cell to the centre of the remap zone.
For this method to correctly resolve a spherical outflow of wind the region must be a minimum of 3 cells in diameter, with all cells at the same refinement level.
In the case of all of our simulations we have utilised stars with a minimum of 6 cells in diameter.
In order to assure that the simulation behaves correctly, it was determined experimentally that the remap cells do not intersect with the WCR.
Orbits are calculated by moving the remap zones in a manner consistent with Keplerian dynamics, which are repositioned at the start of every timestep.
This orbital speed is also added to the remap wind speed.

\subsection{Gas and dust cooling}
\label{sec:gas-dust-cooling}

\begin{figure}
  \centering
  \includegraphics[width=\linewidth]{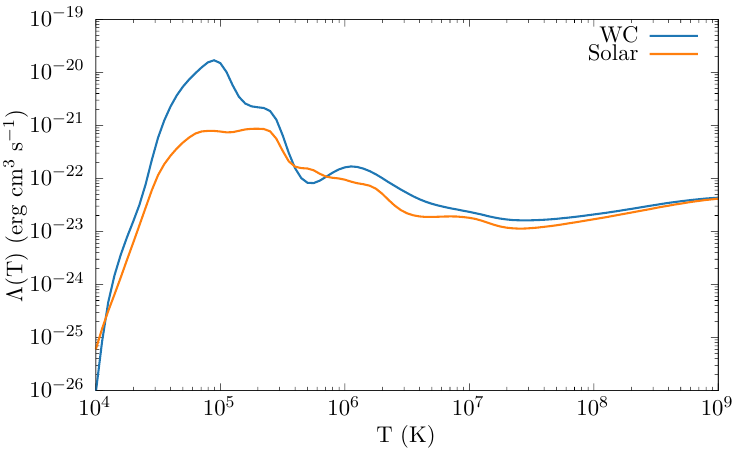}
  \caption[WR and OB $\Lambda(T)$ cooling curves]{Comparison of WC and solar cooling curves for calculating the energy loss due to gas cooling.}
  \label{fig:cooling-curve}
\end{figure}

Cooling due to photon emission from atoms, ions and free electrons, as well as dust particles, is simulated by removing energy from the cells at each timestep.
The total energy loss is calculated by integrating the energy loss rates due to gas, plasma and dust cooling using the Euler method; in regions with very rapid cooling sub-stepping is used to improve accuracy, with the number of sub-steps being determined by comparing the timestep to the cooling timescale of the cell.
Gas cooling is simulated using a lookup table method.
A data file containing the gas temperature and associated normalised emissivity, $\Lambda\rms{w}(T)$, of the wind at that temperature is read into the simulation.
In a typical cooling step, the temperature is calculated and compared with the lookup table to find the closest temperature bins that are lower and higher than the cell temperature.
A linear interpolation is then performed to find an appropriate value for $\Lambda\rms{w} (T)$.
As the two winds have significantly different abundances and can be thoroughly mixed in the WCR, we calculate an emissivity value for gas in a particular cell with the equation

\begin{equation}
  \Lambda\rms{g}(T) = C\Lambda\rms{w,WR}(T) + (1-C)\Lambda\rms{w,OB}(T) , 
\end{equation}

\noindent
where $C$ is the wind ``colour'', or mixing fraction, where 1 is a pure WR wind and 0 is a pure OB wind.
The rate of change in energy per unit volume due to plasma and gas cooling, $\mathcal{L}\rms{g}$, is then calculated through the equation:

\begin{equation}
  \mathcal{L}\rms{g} = \left(\frac{\rho\rms{g}}{m\rms{H}}\right)^2 \Lambda\rms{g}(T),
\end{equation}

\noindent
where $\rho\rms{g}$ is the gas density and $m\rms{H}$ is the mass of a hydrogen atom.
The lookup table was generated by mixing a series of cooling curves generated by MEKAL simulations of elemental gasses.
These simulations were combined based on the elemental abundances of each wind, with the WC star having typical WC9 abundances and the OB star having a solar abundance (see Table \ref{tab:abundances}).
Figure \ref{fig:cooling-curve} shows the resulting cooling curves used for each star.
The most significant abundances used are noted in Table \ref{tab:abundances}.
The cooling regime of the simulations ranges between temperatures of $10^4$ to $10^9\,\si{\kelvin}$.
A floor temperature of $10^4$ \si{\kelvin} is implemented.
Temperatures between $\SI{1e4}{\kelvin} < T \leq \SI{1.1e4}{\kelvin}$ are set to $10^4\,\si{\kelvin}$ as they are assumed to be either rapidly cooling or a part of the stellar wind.

\begin{table}
  \centering
  \begin{tabular}{@{}lll@{}}
  \toprule
  \multicolumn{1}{l}{} & \multicolumn{2}{c}{X(E)} \\ \cmidrule(l){2-3} 
   & Solar & WC9 \\ \midrule
  H & $0.705$ & $0.0$ \\
  He & $0.275$ & $0.546$ \\
  C & $3.07 \times 10^{-3}$ & $0.4$ \\
  N & $1.11 \times 10^{-3}$ & $0.0$ \\
  O & $9.60 \times 10^{-3}$ & $0.05$ \\
  \hline
  \end{tabular}
  \caption[Abundances by mass used for OB and WR stars]{Abundances by mass used for the OB and WR stars being simulated. Other elements are assumed trace when calculating dust emission \citep{williamsSpectraWC9Stars2015}.}
  \label{tab:abundances}
\end{table}

A model for cooling due to emission from dust grains is also included as dust cooling is expected to play a significant role in each system.
The rate of cooling is calculated using the uncharged grain case of the \cite{dwek_infrared_1981} prescription.
Grains are heated due to collisions with ions and electrons, causing them to radiate, with energy being removed from the simulation.
This assumes that the region being simulated is optically thin to far infrared photons.
The grain heating rate (in \si{\erg\per\second}) is calculated with the following formula

\begin{equation}
    H = 1.26 \times 10^{-19} \frac{n}{A^{1/2}} a^2(\si{\micro\metre}) T^{3/2} h(a,T) , 
\end{equation}

\noindent
where $H$ is the heating rate due to atom and ion collisions, 
$n$ is the particle number density,
$A$ is the mass of the incident particle in AMU,
$a(\si{\micro\metre})$ is the grain radius in microns,
$T$ is the temperature of the ambient gas,
and $h(a,T)$ is the effective grain ``heating factor'', also referred to as the grain transparency. 

To obtain the collisional heating due to incident atoms, $H_\text{coll}$, the heating rates are summed for hydrogen, helium, carbon, nitrogen and oxygen atom collisions:

\begin{equation}
  H_\text{coll} = H_\text{H} + H_\text{He} + H_\text{C} + H_\text{N} + H_\text{O} .
\end{equation}

\noindent
Other elements are not considered as they are present in trivial proportions in both winds.
As dust grains are assumed to be uncharged, the grain transparency for each species is calculated with the formula

\begin{equation}
  h(a,T) = 1 - \left( 1 + \frac{E_0}{2 k_\text{B} T} \right) e^{- E_0 / k_\text{B} T} ,
\end{equation}

\noindent
where $E_0$ is the initial energy required to overcome the grain's potential and $k_\text{B}$ is the Boltzmann constant.

\begin{figure}
  \centering
  \includegraphics[width=\linewidth]{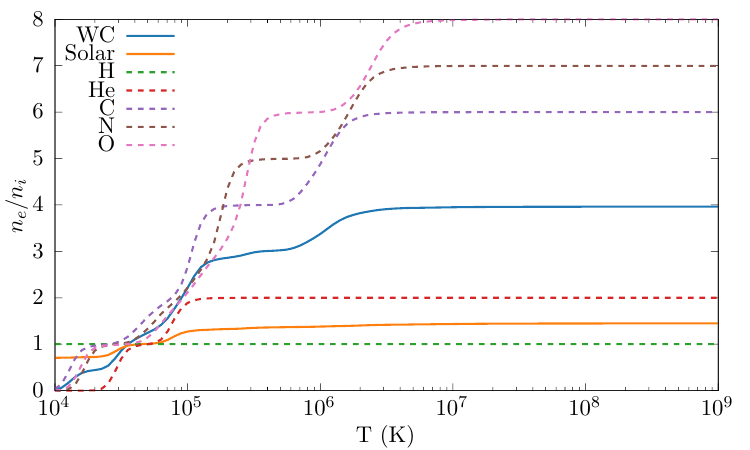}
  \caption[OB and WR electron-ion ratios]{A comparison of the electron-ion ratio in both winds as a function of temperature. Also shown are the electron-to-ion ratios for the individual elements.}
  \label{fig:electron-curve}
\end{figure}

Electron-grain collisional heating, $H_\text{el}$, is modelled using the same calculation for $H_\text{coll}$, albeit with some differences.
For accurately calculating the energy loss due to electron collisions, the electron number density, $n\rms{e}$, needs to be known.
This is achieved with a second series of lookup tables that contain the electron-to-ion ratio of each wind across a temperature range of $10^4$ to $10^9\,\si{\kelvin}$ (Fig. \ref{fig:electron-curve}).
The electron number density is $n \rms e = \beta n \rms i$, where $\beta$ is the electron-to-ion ratio and $n \rms i$ is the ion number density.
Another difference between calculating electron-grain and gas-grain cooling is calculating electron-grain transparency, which is a significantly more complex problem than calculating ion-grain transparency.
An assumed full opacity proves to be extremely inaccurate at temperatures $>10^6\,\si{\kelvin}$.
Electron-grain transparency is therefore calculated via an approximation described in \cite{dwek_infrared_1981}:

\begin{equation}
  \begin{alignedat}{3}
    h(x^*) & = 1 ,                && ~~ x^* > 4.5, \\
           & = 0.37{x^*}^{0.62} , && ~~ x^* > 1.5 , \\
           & = 0.27{x^*}^{1.50} , && ~~ \text{otherwise,}
  \end{alignedat}
\end{equation}

\noindent
where $x^* = \num{2.71e8} a^{2/3} (\si{\micro\metre})/T$.
This approximation is approximately 4 orders of magnitude faster than using an integration method, while differing by less than 8\%.
Grain-grain collisions are not modelled, as this would be difficult to calculate due to the single-fluid model in use.
Further simulations utilising a multi-fluid model could allow for this to be simulated.
Finally, in order to calculate the change in energy due to dust cooling, the rate of energy change, $\mathcal{L}\rms{d}$, is calculated using the formulae:

\begin{subequations}
  \label{eq:transparency-approximation}
  \begin{align}
    H\rms{T}           & = H_\text{coll} + H_\text{el} , \\
    \mathcal{L}\rms{d} & = n_\text{d} H\rms{T} ,
  \end{align}
\end{subequations}

\noindent
where $H\rms{T}$ is the total grain heating rate
and $n_\text{d}$ is the dust number density.
The total energy loss rate per unit volume due to gas and dust cooling is given by:

\begin{equation}
	\mathcal{L}\rms{T} = \mathcal{L}\rms{g} + \mathcal{L}\rms{d} .
\end{equation}

\subsection{Numerical modelling of dust through advected scalars}

The most important modification to Athena++ was the addition of a dust growth and destruction model to simulate the production of dust within the WCR.
A series of passive scalars were used where the dust parameters described by the scalars can evolve and advect through the simulation, analogous to a co-moving fluid, which previous work has noted is an accurate dynamical model for dust within the WCR \citep{hendrix_pinwheels_2016}.
In these simulations, information about the dust is stored in the form of two variables, the average grain radius, $a$, and the dust-to-gas mass ratio, $z$.
From these constants the dust growth rate, number density, and total dust mass can be derived.
A co-moving model allows for a simplified model of dust formation. In such a model, the mean relative velocity between the dust and gas is given by:

\begin{equation}
  \langle u \rangle = \sqrt{\frac{8kT}{\pi m \rms r}} ,
\end{equation}

\noindent
where $m \rms r$ is the familiar reduced mass between a test particle of mass $m \rms t$ and a field particle of mass $m \rms f$, such that $m \rms r = m \rms f m \rms t / (m \rms f + m \rms t)$.

\noindent
As the dust grain is significantly more massive, the reduced mass is approximately equal to the grain mass, simplifying the dynamics of the simulation in a co-moving case.
Throughout our simulations we observed a dust grain size that suggests the grains should be dynamically coupled to the gas, though turbulent mixing with larger dust grains would present an interesting avenue of future research.
Furthermore, the effect of radiation pressure on the dust grains is not simulated, as it was assumed that the dust grains would be sufficiently shielded from the radiation from their parent stars.

In this model, growth and destruction occur in distinct temperature regimes.
Dust growth occurs when $T \leq \SI{1.4e4}{\kelvin}$ whilst dust destruction occurs at temperatures of $T \geq 10^6 \, \si{\kelvin}$.

Dust growth is modelled through approximating growth due to grain-gas accretion where grains co-moving with a gas perform relatively low-velocity collisions with the surrounding gas, causing it to accrete onto the surface of the dust grain 
\citep{spitzer_jr._physical_2008}.
Assuming a single average grain size the rate of change in the average grain radius is given by:

\begin{equation}
  \frac{da}{dt} = \frac{\xi \rho\rms{C} w_\text{C}}{4 \rho\rms{gr}} ,
\end{equation}

\noindent
where $w\rms{C}$ is the Maxwell-Boltzmann distribution RMS velocity for carbon ($w\rms{C} = \sqrt{3 k\rms{B} T/12 m\rms{H}}$), $\xi$ is the grain sticking efficiency, $\rho\rms{C}$ is the carbon density in the gas ($\rho\rms{C} = X(C)\rho\rms{g}$, where $X(C)$ is the wind carbon abundance), and $\rho_\text{gr}$ is the grain bulk density.
The associated rate of dust density change, $d\rho\rms{d}/dt$ is calculated with the formulae:

\begin{subequations}
  \begin{align}
    \frac{dm\rms{gr}}{dt} & = 4\pi \rho\rms{gr} \frac{da}{dt} a^2 = \pi \xi \rho\rms{C} w\rms{C} a^2, \\
    \frac{d\rho\rms{d}}{dt} & = \frac{dm\rms{gr}}{dt} n\rms{d} ,
  \end{align}
\end{subequations}

\noindent
where $n_\text{d}$ is the grain number density and $dm\rms{gr}/dt$ is the rate of change of the grain mass.
While $\xi \rightarrow 1$ in the case of cold, neutral gas \citep{spitzer_jr._physical_2008}, the sticking parameter for carbon grains decreases significantly
as the gas temperature increases \citep{devlinGrainSticking}.
As such, we take $\xi = 0.1$ as a conservative value throughout this paper. Furthermore, high values of $\xi$ were found to cause significant simulation instability.
A bulk density analogous to amorphous carbon grains ($\rho\rms{gr} = \SI{3.0}{\gram\per\centi\metre\cubed}$) is used.

Dust destruction gas-grain sputtering is calculated using the \cite{drainePhysicsDustGrains1979} prescription.
A dust grain has a lifetime which is dependent on the number density of the gas the grain is moving through, $n\rms{g}$.
In the case of amorphous carbon grains, the dust lifetime is:

\begin{equation}
  \tau\rms{gr} = \frac{a}{da/dt} \approx \SI{3e6}{\year} \cdot \frac{a (\si{\micro\metre})}{n\rms{g}} \equiv \num{9.467e17} \cdot \frac{a}{n\rms{g}} .
\end{equation}

\noindent
This value is based on an average lifetime of carbon grains in interstellar shocks at shock temperatures between $10^6$ and $\SI{3e8}{\kelvin}$ \citep{tielens_physics_1994,dwekCoolingSputteringInfrared1996}.
The rate of change in grain radius can be calculated with the formula

\begin{equation}
  \frac{da}{dt} = - \frac{a}{\tau\rms{gr}} = - \num{1.056e-18} \cdot n\rms{g},
\end{equation}

\noindent
The rate of change in grain mass and dust density can then be calculated with the formulae:

\begin{subequations}
  \begin{align}
    \frac{d m\rms{gr}}{dt} & = 4 \pi \rho\rms{gr} a^2 \frac{da}{dt} = \num{-1.33e-17} \cdot n\rms{g} \rho\rms{gr} a^2 , \\
    \frac{d \rho\rms{d}}{dt} & = \frac{d m\rms{gr}}{dt} n\rms{d}.
  \end{align}
\end{subequations}

\noindent
Application of the dust growth and destruction routines in the code is determined by the gas temperature of a cell.

In order to propagate dust through each simulation, a small initial value for the advected scalars is set in each cell in the remap zones.
An initial grain radius of $a_i = 50 \, \text{\AA}$ and initial dust-to-gas mass ratio of $z_i = 10^{-6}$ is imposed.
Changing $z_i$ does not significantly impact the final dust-to-gas mass ratio of the system as $z$ rapidly increases within the WCR and dust growth in the WCR dominates the total production.
Dust also grows to some extent in the unshocked winds but at a much lower rate than within the WCR.
A small initial grain radius is sensible, as small dust grains are believed to rapidly nucleate from impinging carbon ions 
\citep{harriesThreedimensionalDustRadiativetransfer2004,zubkoPhysicalModelDust1998a}.

In order to determine if our dust model is producing reasonable dust yields, we calculate the maximum expected dust production rate in each system, $\dot{\text{M}}_\text{d,max}$.
This rate would occur if 100\% of the carbon in the WR wind being shocked by the WCR was converted into dust.
The fraction of the WR wind that passes through the WCR is given by

\begin{equation}
	f_\text{WR} = \frac{1 - \cos \left(\theta_\text{WR}\right)}{2} ,
\end{equation}

\noindent
where $\theta_\text{WR}$ is the opening angle of the WR shock front, approximated as $\theta_\text{WR} \approx 2 \tan^{-1} ( \eta^{1/3} ) + \pi/9$ \citep{pittardCollidingStellarWinds2018}.
The theoretical maximum dust production rate is then

\begin{equation}
  \label{eq:maxdust}
	\maxdust = \dot{\text{M}}_\text{WR} \text X_\text{C,WR} f_\text{WR},
\end{equation}

\noindent
where $\text X_\text{C}$ is the carbon mass fraction in the WR star.
The effect of carbon depletion is not simulated as only extremely high dust conversion rates would significantly impact the abundance of carbon in a WC wind.
While some simulations produced values of $z$ on the order of 10\%, this would still only slightly decrease the amount of carbon in the wind.
However, in the case of systems with an extremely high dust production rate such as WR104 carbon depletion would need to be correctly simulated.

\section{Model Parameters}
\label{sec:p1-model-parameters}

In this paper we do not attempt to model particular systems.
Rather we aim to gain a deeper understanding of the primary influences of dust formation in a CWB system.
A series of simulations were therefore run in order to determine how dust formation varies due to changes in orbital separation and wind momentum ratio.
A baseline simulation with properties similar to WR98a with a circular orbit and identical stellar masses was created.
This baseline simulation has a momentum ratio of $0.02$.
Other simulations were then run with different orbital separations and/or wind momentum ratios.
Another set of simulations were run where the cooling mechanisms were selectively disabled, in order to understand how radiative cooling affects the dust production rate.
Tables \ref{tab:baseline-windproperties} and \ref{tab:baseline-orbits} detail the wind and orbital parameters of the baseline simulation.
The orbital separation is modified by changing the orbital period of the simulation, while the wind momentum ratio is modified by adjusting the mass loss rate and wind terminal velocity for each star.
Two simulation sub-sets for this were performed: simulations where the wind terminal velocities were adjusted for each star and simulations where the mass loss rates for each star were adjusted.

\begin{table}
  \centering
  \begin{tabular}{lll}
  \hline
  Parameter & WR & OB \\ \hline
  $\dot M$ & \SI{5.0e-6}{\solarmass\per\year} & \SI{5.0e-8}{\solarmass\per\year} \\
  $v_\infty$ & \SI{1.0e8}{cm.s^{-1}} & \SI{2.0e8}{cm.s^{-1}} \\
  $T_w$ & \SI{1.0e4}{\kelvin} & \SI{1.0e4}{\kelvin} \\
  \hline
  \end{tabular}
  \caption{Wind properties of the baseline system.}
  \label{tab:baseline-windproperties}
\end{table}

\begin{table}
  \centering
  \begin{tabular}{ll}
  \hline
  Parameter & Value \\
  \hline
  $\text{M}_\text{WR/OB}$ & 10.0 \si{\solarmass} \\
  $d_\text{sep}$ & \SI{4.0}{\au} \\
  $P$ & \SI{1.80}{\year} \\
  \hline
  \end{tabular}
  \caption{Baseline system orbital properties.}
  \label{tab:baseline-orbits}
\end{table}

\subsection{Cooling mechanisms}

For this set of simulations, the influence of cooling was changed by varying which cooling routines are operating.
All simulations in this set keep the same orbital and wind parameters, which are that of the baseline system described in Tables \ref{tab:baseline-windproperties} \& \ref{tab:baseline-orbits}.
One simulation has both plasma and dust cooling in operation (the \texttt{fullcool} simulation), while the other two simulations have plasma cooling only and no cooling, respectively (\texttt{plasmacool} and \texttt{nocool}, Table \ref{tab:cooling-param}).
The final, no radiative cooling simulation instead relies on adiabatic expansion for temperature change in the WCR; as such, this simulation behaves as if it has a $\chi$ value for both winds that is arbitrarily high.
The post-shock flow in the \texttt{nocool} model will also be unable to compress as much due to the lack of energy loss via radiative cooling.
The role of these simulations is to discern whether cooling alone, or other system parameters can affect dust production.

\begin{table}
  \centering
  \begin{tabular}{lll}
    \hline
    Name & Plasma cooling? & Dust cooling? \\
    \hline
    \texttt{fullcool} & Yes & Yes \\ 
    \texttt{plasmacool} & Yes & No \\
    \texttt{nocool} & No & No \\
    \hline
  \end{tabular}
  \caption{Cooling series simulation parameters.}
  \label{tab:cooling-param}
\end{table}

\subsection{Wind momentum ratio}

Another set of simulations was devised in order to assess the influence of the wind parameters on the formation of dust within a CWB.
As the wind momentum ratio is dependent on both the mass loss rate and wind velocity of each star, each of these properties is modified over a set of different simulations.
$\eta$ is varied from 0.01 to 0.04 by adjusting the wind parameters for each star.
This is further subdivided by which property is modified, either the mass loss rate or wind terminal velocity (Table \ref{tab:vinf-param}).
As the cooling parameter, $\chi$, has a much stronger dependency on $v^\infty$ than $\dot{\text{M}}$, the modification of either parameter while maintaining a similar value for $\eta$ allows us to determine whether $\chi$ is the primary parameter determining the formation of dust within WCd systems.
This can be seen when comparing simulations \texttt{mdot-1} and \texttt{vinf-1}, which have similar wind momentum ratios but the cooling parameters for the WC star differ by a factor of 32.
These simulations are compared to the baseline simulation, which has a radiative post-shock WCR.
All simulations were run for a minimum of 1 orbit.
As these orbits are circular, there should be no major variance of the winds after the start-up transients are fully advected, save for some fluctuations.

\begin{table*}
  \centering
  \begin{tabular}{llllllll}
  \hline
  Name & $\dot{\text{M}}_\text{WR}$ & $\dot{\text{M}}_\text{OB}$ & $v^\infty_\text{WR}$ & $v^\infty_\text{OB}$ & $\eta$ & $\chi_\text{WR}$ & $\chi_\text{OB}$ \\ 
  \hline
  \texttt{baseline}& \SI{5.0e-6}{\solarmass\per\year} & \SI{5.0e-8}{\solarmass\per\year} & \SI{1e8}{cm.s^{-1}} & \SI{2e8}{cm.s^{-1}} & 0.02 & 1.20 & 1915 \\
  \texttt{mdot-1}  & \SI{1.0e-5}{\solarmass\per\year} & \SI{5.0e-8}{\solarmass\per\year} & \SI{1e8}{cm.s^{-1}} & \SI{2e8}{cm.s^{-1}} & 0.01 & 0.60 & 1915 \\
  \texttt{mdot-2}  & \SI{2.5e-6}{\solarmass\per\year} & \SI{5.0e-8}{\solarmass\per\year} & \SI{1e8}{cm.s^{-1}} & \SI{2e8}{cm.s^{-1}} & 0.04 & 2.39 & 1915 \\
  \texttt{mdot-3}  & \SI{5.0e-6}{\solarmass\per\year} & \SI{1.0e-7}{\solarmass\per\year} & \SI{1e8}{cm.s^{-1}} & \SI{2e8}{cm.s^{-1}} & 0.04 & 1.20 & 957  \\
  \texttt{mdot-4}  & \SI{5.0e-6}{\solarmass\per\year} & \SI{2.5e-8}{\solarmass\per\year} & \SI{1e8}{cm.s^{-1}} & \SI{2e8}{cm.s^{-1}} & 0.01 & 1.20 & 3830 \\
  \texttt{vinf-1}   & \SI{5.0e-6}{\solarmass\per\year} & \SI{5.0e-8}{\solarmass\per\year} & \SI{2e8}{cm.s^{-1}} & \SI{2e8}{cm.s^{-1}} & 0.01 & 19.1 & 1915  \\
  \texttt{vinf-2}   & \SI{5.0e-6}{\solarmass\per\year} & \SI{5.0e-8}{\solarmass\per\year} & \SI{5e7}{cm.s^{-1}} & \SI{2e8}{cm.s^{-1}} & 0.04 & 0.07 & 1915  \\
  \texttt{vinf-3}   & \SI{5.0e-6}{\solarmass\per\year} & \SI{5.0e-8}{\solarmass\per\year} & \SI{1e8}{cm.s^{-1}} & \SI{4e8}{cm.s^{-1}} & 0.04 & 1.20 & 30638 \\
  \texttt{vinf-4}   & \SI{5.0e-6}{\solarmass\per\year} & \SI{5.0e-8}{\solarmass\per\year} & \SI{1e8}{cm.s^{-1}} & \SI{1e8}{cm.s^{-1}} & 0.01 & 1.20 & 120   \\
  \hline
  \end{tabular}
  \caption[Terminal velocity series wind parameters]{Wind parameters for simulations varying the wind mass loss rate, $\dot{\text{M}}$, and terminal velocity, $v^\infty$. $\eta$ is the wind momentum ratio (Eq. \ref{eq:paper1-eta}), and $\chi$ is the cooling parameter (Eq. \ref{eq:paper1-chi}). Note that the value of $\chi$ does not take into account cooling due to dust.}
  \label{tab:vinf-param}
\end{table*}

\subsection{Separation distance}

A final series of simulations was performed with the wind parameters equivalent to the baseline model, but with differing orbital separations.
The separation was altered by modifying the orbital period.
The separation distance was varied from the baseline model of \SI{4}{\au} up to \SI{64}{\au} (Table \ref{tab:dsep-param}), which has the effect of modifying the cooling parameter, $\chi$, of each simulation without changing the wind momentum ratio; allowing us to further discern which is the dominant parameter influencing dust formation.
For instance, simulation \texttt{dsep-64AU} has a cooling parameter value approaching the fast WR wind model \texttt{vinf-1}, despite having a wind momentum ratio of 0.02.

Each simulation has a coarse resolution of $320 \times 320 \times 40$ cells, with a varying number of levels.
As the separation distance is doubled, the static mesh refinement box around the stars is doubled in size and the number of levels is decremented. This manipulation of levels ensures that the number of cells between the stars is kept consistent and reduces memory usage.
The extent for all simulations in this series were doubled over the other series in this paper to approximately $2000 \times 2000 \times 250 \, \si{\au}$.
Similarly to the previous set of simulations, a minimum of 1 orbit was needed for each simulation, however, as the orbital period of each simulation varies, certain simulations were able to run for a significantly longer length of time, with data for multiple orbits being obtained.

\begin{table*}
  \centering
  \begin{tabular}{lllllll}
    \hline
    Name & P & $d_\text{sep}$ & $\chi_\text{WR}$ & $\chi_\text{OB}$ & Levels & Effective Resolution \\
	\hline
    \texttt{dsep-4AU}  & \SI{1.80}{\year} & \SI{4}{\au}  & 1.20 & 1915  & 7 & $(20480 \times 20480 \times 2560) \,\text{cells}$ \\
    \texttt{dsep-8AU}  & \SI{5.06}{\year} & \SI{8}{\au}  & 2.39 & 3830  & 6 & $(10240 \times 10240 \times 1280) \,\text{cells}$ \\
    \texttt{dsep-16AU} & \SI{14.3}{\year} & \SI{16}{\au} & 4.79 & 7659  & 5 & $(5120 \times 5120 \times 640) \,\text{cells}$    \\
    \texttt{dsep-32AU} & \SI{40.5}{\year} & \SI{32}{\au} & 9.57 & 15319 & 4 & $(2560 \times 2560 \times 320) \,\text{cells}$    \\
    \texttt{dsep-64AU} & \SI{115}{\year}  & \SI{64}{\au} & 19.1 & 30638 & 3 & $(1280 \times 1280 \times 160) \,\text{cells}$    \\ \hline
  \end{tabular}
  \caption{Parameters of simulations varying the separation distance, $d\rms{sep}$, between the stars.}
  \label{tab:dsep-param}
\end{table*}

\subsection{Data collection}

HDF5 files were generated at regular time intervals containing the primitive variables of the simulation: gas density, $\rho$, gas pressure, $P$, and wind velocity components, $v_x$, $v_y$ and $v_z$.
These variables were then used to derive other variables such as temperature and energy.
The scalars containing the dust-to-gas mass ratio, $z$, is also included.
The wind ``colour'', the proportion of gas from each star, was also tracked.
A value of 1.0 indicates a pure WR wind while 0.0 indicates a pure OB wind.

The volume-weighted totals of all parameters of interest were also collected, such as the total gas and dust mass of the system and average grain radius.
Average values, such as $\bar{z}$ and $\bar{a}$, are mass-weighted.
To calculate dust formation within the WCR, a method of determining if a cell was part of the wind collision region was devised - the cell density would be compared to the predicted density of a single smooth wind with the wind parameters of the WC star in the system:

\begin{equation}
  \rho_\text{WC} = \frac{\dot{M}_{WC}}{4 \pi r^2 v^\infty_{WC}},
\end{equation}

\noindent
where $r$ is the distance from the barycentre.
This threshold value was set to $1.25\rho_\text{SW}$.
Higher threshold values were found to be inaccurate at large distances from the barycentre.
Other methods of detecting the WCR, such as determining wind mixing levels, were not successful in general.

\section{Results}
\label{sec:p1-results}

\begin{figure}
  \centering
  \includegraphics[width=\linewidth]{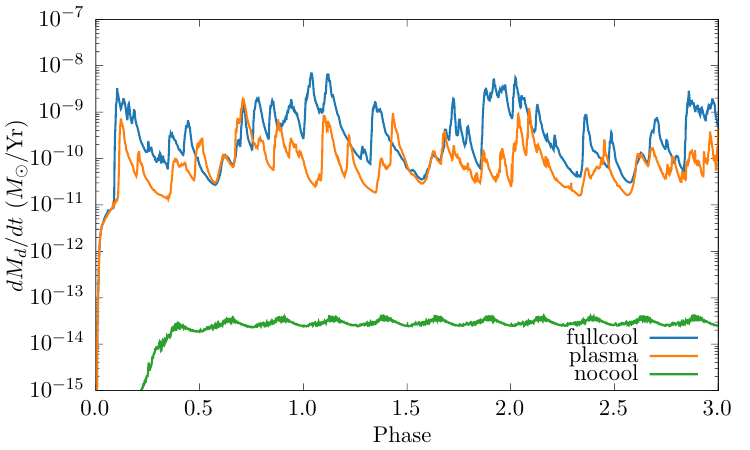}
  \caption[Comparison of dust formation rates with cooling methods]{A comparison of the dust formation rates as the cooling mechanisms in the simulation are changed. Without adequate cooling barely any dust is formed. While dust formation increases with all cooling mechanisms enabled, plasma cooling is still the dominant cooling process between $10^4$ and $10^9 \, \si{\kelvin}$ for dust production.}
  \label{fig:coolingprocess-dustproduction}
\end{figure}

\begin{figure}
  \centering
  \includegraphics[width=\linewidth]{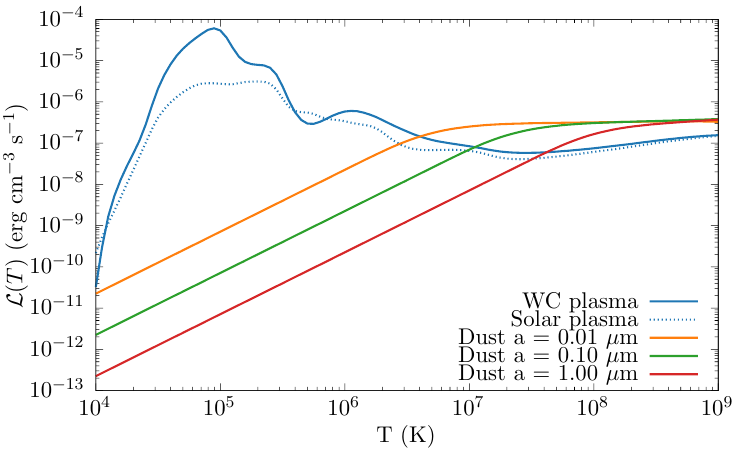}
  \caption[Comparison of dust and plasma cooling rates in post-shock environment]{Comparison of the energy loss rate due to plasma and dust cooling with varying grain sizes, where $\rho_g = 10^{-16} \, \si{\gram\per\centi\metre\cubed}$ (typical of the density in the WCR) and a dust-to-gas mass ratio of $10^{-4}$ is assumed. Whilst less influential at lower temperatures, dust cooling can aid the overall cooling in the immediate high temperature post-shock environment.}
  \label{fig:postshockcoolcomparison}
\end{figure}

The first set of simulations were performed in order to assess whether the implemented cooling model would influence dust formation within the WCR.
This was found to be the case.
Figure \ref{fig:coolingprocess-dustproduction} shows that with no cooling only a very small amount of dust formation occurs.
Dust production in the radiative simulations is significantly higher, with the \texttt{fullcool} simulation having consistently higher dust formation rates than the \texttt{plasmacool} simulation.
Figure \ref{fig:postshockcoolcomparison} shows that at the temperatures present within the WCR, dust grains that are present can enhance the cooling, allowing the shocked gas to reach temperatures low enough for dust formation faster than if only plasma cooling was simulated.

\begin{figure*}
  \centering
  \includegraphics{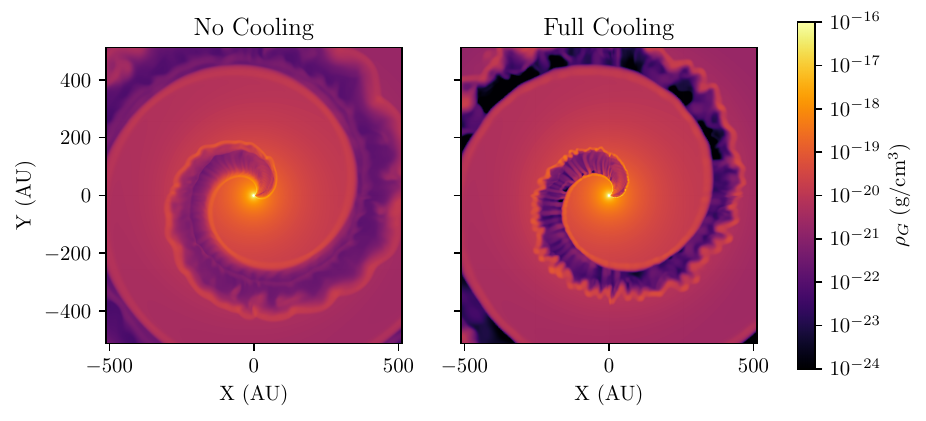}
  \caption[Instabilities due to cooling]{Density comparison in the orbital plane for the \texttt{nocool} and \texttt{fullcool} models. With cooling enabled instabilities are far more prevalent, with pockets of very high density material within the WCR.}
  \label{fig:coolingprocess-density}
\end{figure*}

In the case of the \texttt{fullcool} simulation, a peak dust formation rate of $\SI{7e-09}{\solarmass\per\year}$ was calculated.
This fluctuation appears to be due to dust forming mostly in high density clumps (see Fig. \ref{fig:coolingprocess-density}).
The average dust formation rate from these simulations is noted in Table \ref{tab:radiative-average-rates}.
The observed rates are less than 0.1\% of the theoretical maximum given by Eq. \ref{eq:maxdust}, which indicates that the average dust-to-gas ratio, $z$, in the WCR, does not exceed $10^{-3}$.

\begin{figure*}
  \centering
  \includegraphics{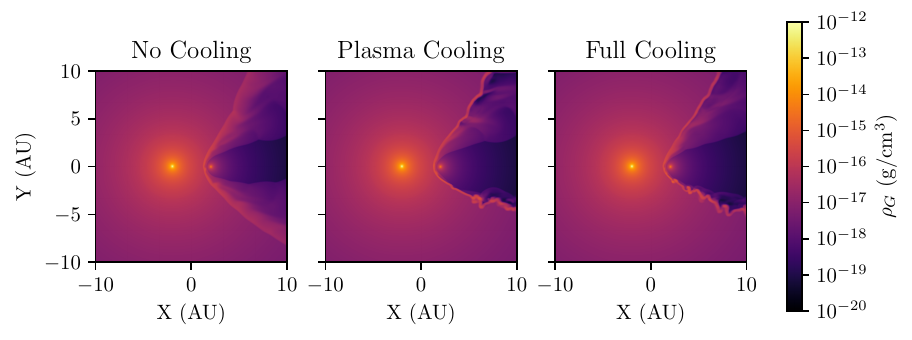}
  \caption[Density comparison of simulations with differing radiative processes]{Density comparison  in the orbital plane of simulations with differing radiative processes.}
  \label{fig:postshockcompression}
\end{figure*}

\begin{figure*}
  \centering
  \includegraphics{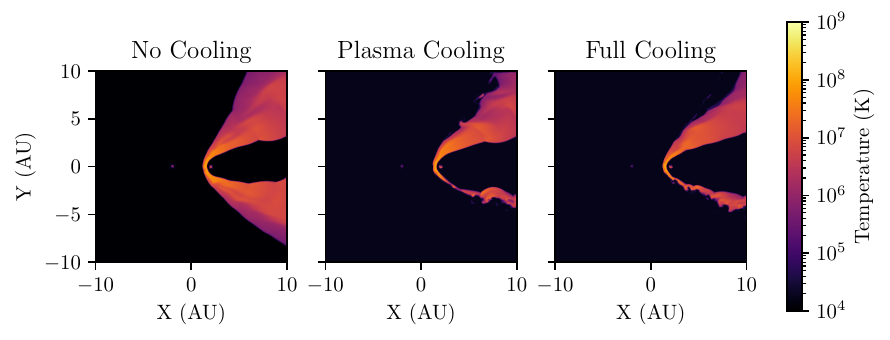}
  \caption[Temperature comparison of simulations with differing radiative processes]{Temperature comparison in the orbital plane of simulations with differing radiative processes.}
  \label{fig:postshocktemperature}
\end{figure*}

\begin{figure}
  \centering
  \includegraphics[width=\linewidth]{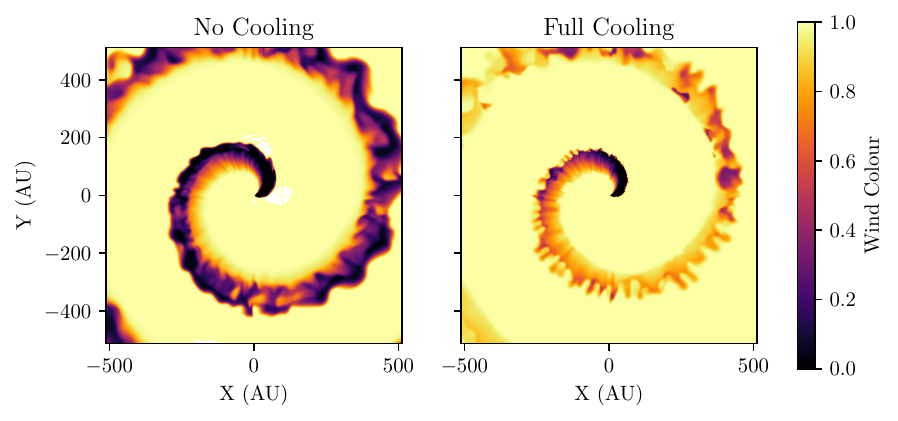}
  \caption[Wind mixing due to radiative methods]{Wind ``colour'' for \texttt{nocool} and \texttt{fullcool} models. The WCR is more thoroughly mixed if the simulation is allowed to cool.}
  \label{fig:radiative-windmixing}
\end{figure}

\begin{figure}
  \centering
  \includegraphics[width=\linewidth]{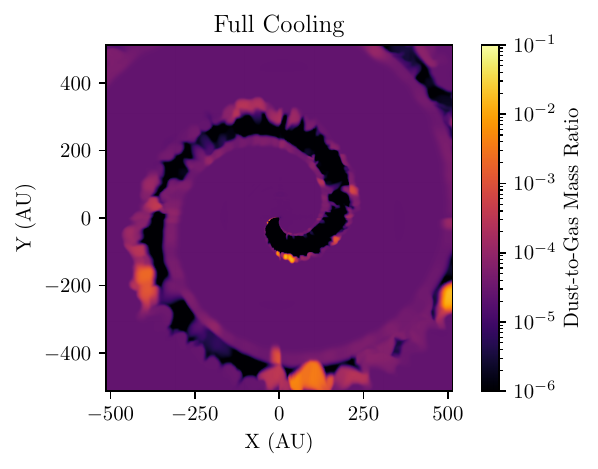}
  \caption[\texttt{Baseline} simulation $z$, full extent]{The full extent of the \texttt{baseline} simulation, showing the dust-to-gas mass ratio in the orbital plane. Dust typically forms in clumps within instabilities, leading to a variation of the dust formation rate as the simulation progresses. Most of the dust forms in the leading arm of the WCR.}
  \label{fig:full-radiative-z}
\end{figure}

As cooling is significant in the post-shock WR wind ($\chi_\text{WR} = 1.2$), further compression occurs, resulting in much higher post-shock densities (Fig. \ref{fig:postshockcompression}).
This rapid cooling results in ideal conditions for dust formation, especially within high density instabilities.
A similar effect for the OB wind is not observed, as radiative energy losses are not influential on the dynamics of the flow, due to the faster, significantly less dense stellar wind ($\chi_\text{OB} = 1915$).
Fig. \ref{fig:postshocktemperature} shows that the \texttt{fullcool} simulation has a similar immediate-post shock temperature to an adiabatic model, but the shocked WR wind cools to the floor temperature within an extremely short timescale, allowing the nascent dust grains to grow.
We also observe that simulations with cooling have a markedly more mixed wind, due to instabilities in the post-shock environment (Fig. \ref{fig:radiative-windmixing}).
Fig. \ref{fig:full-radiative-z} shows that dust clumps form shortly after the initial wind collision.
These clumps rapidly convert post-shock gas to dust.
However, rapid dust production tapers off as the post-shock flow becomes more diffuse.
This behaviour is similar to the dust simulations described in \cite{harriesThreedimensionalDustRadiativetransfer2004} and \cite{hendrix_pinwheels_2016}, which indicate that the bulk of dust formation occurs only a short distance from the parent stars.
The post-shock temperature is significantly lower in the leading edge of the WCR relative to the orbital motion, leading to a larger portion of dust forming in this region.
The lower temperature is due to the more rapid cooling caused by the higher density on this side of the WCR (see below).

\cite{pittard_3d_2009} notes that in the case of colliding winds with $\eta = 1$ the trailing edge of the WCR takes part in oblique shocks with the stellar winds, while the leading edge is shadowed by the upstream WCR from the colliding material.
This results in a trailing edge with strong instabilities and cool, high density clumps of post-shock wind, while the leading edge has a low density flow that is not dominated by instabilities.
This does not appear to occur in these low-$\eta$ systems, as oblique shocks occur at a much greater distance, where the stellar wind is significantly less dense.
Instead, the leading edge of the WCR appears to be much thinner and denser than the trailing edge.
This is believed to be due to the leading edge interacting more strongly with the outflowing material due to the orbital motion of the stars, sweeping up material and obliquely shocking with the downstream WCR. %
Most of the dust formation then occurs in the downstream post-shock region of the leading edge of the WCR, as soon as it has sufficiently cooled (Fig. \ref{fig:full-radiative-z}).
Furthermore, dust formation slows significantly as the post-shock wind begins to diffuse, limiting the dust formation to a region around 100 AU from the WCR apex. %
This is in agreement with \cite{williams_dust_1990} and \cite{hendrix_pinwheels_2016}, who found that there is a limited region suitable for dust formation.

\begin{table}
  \centering
  \begin{tabular}{llllll}
  \hline
  Model & $\eta$ & $\chi_\text{WR}$ & $\chi_\text{OB}$ & $\avgdust$ & $\maxdust$ \\
   &  &  &  & \si{\solarmass\per\year} & \si{\solarmass\per\year} \\ \hline
  \texttt{fullcool} & 0.02   & 1.20 & 1915 & \num{5.38e-10} & \num{9.06E-07} \\ \hline
  \texttt{plasmacool}   & 0.02   & 1.20 & 1915 & \num{1.29e-10} & \num{9.06E-07} \\
  \texttt{nocool}   & 0.02   & 1.20 & 1915 & \num{2.71e-14} & \num{9.06E-07} \\ \hline
  \end{tabular}
  \caption{Average rate of dust production for the set of different radiative simulations. $\maxdust$ is the maximum expected dust formation rate (Eq. \ref{eq:maxdust}).}
  \label{tab:radiative-average-rates}
\end{table}

\subsection{Grain size}

\begin{figure}
  \centering
  \includegraphics[width=\linewidth]{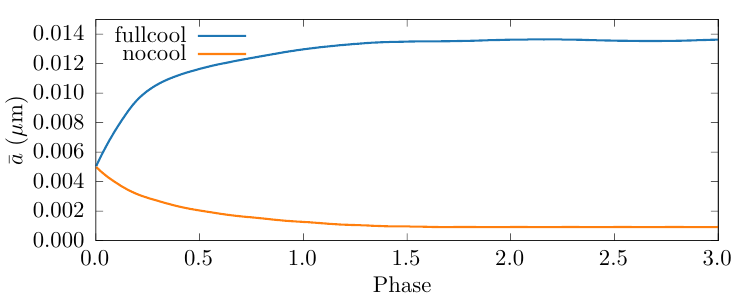}
  \caption[Comparison of grain growth between \texttt{fullcool} and \texttt{nocool} models]{A comparison of mass-averaged grain growth between \texttt{fullcool} and \texttt{nocool} models. Without sufficient cooling dust destruction mechanisms dominate, and rapidly reduce the radius of dust grains in the simulation.}
  \label{fig:rad-grainradius}
\end{figure}

In order to determine the veracity of the dust model, as well as to determine that dust grains are behaving correctly, we observed how dust grains evolve in size over the course of a simulation.
Fig. \ref{fig:rad-grainradius} shows the average grain radius of dust grains within the WCR.
We observe that grain radius steadily grows, before tapering off; this rate of growth is greater in simulations with a greater degree of dust growth.
In the case of the \texttt{nocool} simulation we find that that grain radius decreases; as the WCR cannot cool to temperatures conducive to dust formation, only grain ablation occurs.
This verifies that our model works the other way, and that dust destruction mechanisms can rapidly destroy dust grains if cooling is insufficient in the WCR.
This is expected, as high gas temperatures would readily destroy grains through thermal sputtering.

\begin{figure}
  \centering
  \includegraphics[width=\linewidth]{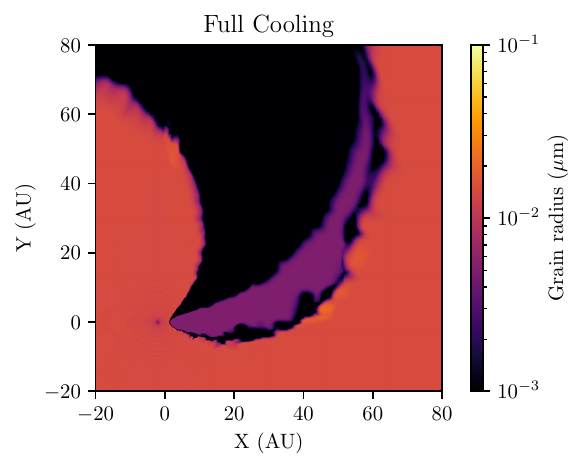}
  \caption[Comparison of grain growth between \texttt{fullcool} and \texttt{nocool} models]{Grain radius plot of the WCR region close to the apex. Grain growth does not occur in the OB wind, and is ablated away in the hot WCR region. Conversely, grain radius growth along the leading edge of the WCR is observed, growing by a factor of 3.}
  \label{fig:rad-r2}
\end{figure}

Fig. \ref{fig:rad-r2} shows the destruction and growth of dust grains in the \texttt{fullcool} model.
We observe that the grain radius has been significantly reduced within the WCR outside of high density clumps where dust production occurs.
We do find some grain growth in regions where dust growth is significant -- in particular along the leading edge of the WCR.
However, this effect is less pronounced than the change in $z$ as seen in Fig. \ref{fig:full-radiative-z}.

\subsection{Mass loss rate variation}

\begin{figure}
  \centering
  \includegraphics[width=\linewidth]{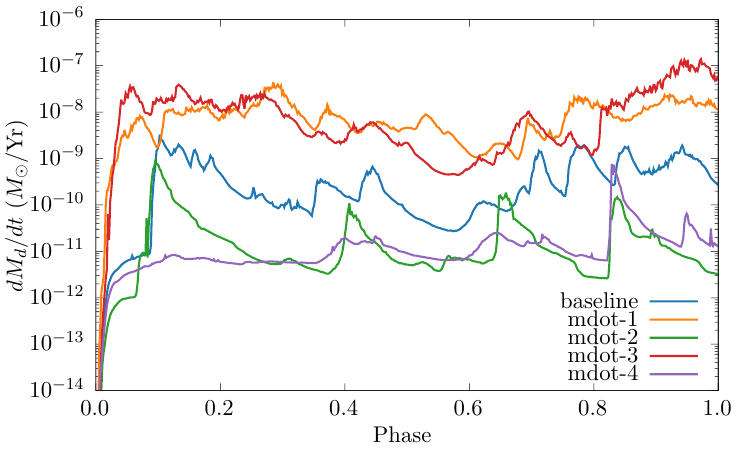}
  \caption[Dust production rate for simulations varying mass loss rate]{A comparison of the dust production rates for simulations that vary the mass loss rate, $\dot M$. Simulations with either a strong primary or secondary wind produce similar levels of dust, whilst if either wind is weaker, the  dust production rate is reduced.}
  \label{fig:mdotdustproductionrate}
\end{figure}

The dust formation rate in the mass loss rate variation simulations was found to be dependent on the strength of the WC or OB winds.
As can be seen in Fig. \ref{fig:mdotdustproductionrate} and Table \ref{tab:mdot-average-rates}, the rates are separated into similar dust production rates for simulations with increases or decreases in mass loss rates; simulations with either wind being stronger than the \texttt{baseline} simulation produced most dust, while simulations with weaker winds produced approximately 3 orders of magnitude less dust than the most productive simulations.
This result appears to be proportional to the wind momentum ratio.
For instance, \texttt{mdot-1} and \texttt{mdot-3} produce on average two orders of magnitude more dust than the \texttt{baseline} simulation.
These simulations have an identical value for $\eta$, but differ in total mass loss rate by a factor of 2.
This suggests that a stronger shock can increase the dust production rate, due to higher post-shock densities and more cooling.
Some of this value can be attributed to the changing number density of grains, particularly in simulations \texttt{mdot-1} and \texttt{mdot-2}, where the relative grain number density increases and decreases by a factor of 2 respectively.
In the case of \texttt{mdot-1} this doubles the number density of grains, increasing the amount of dust cooling, and increases the number of grain nucleation sites for dust formation.

\begin{table}
  \centering
  \begin{tabular}{llllll}
  \hline
  Model & $\eta$ & $\chi_\text{WR}$ & $\chi_\text{OB}$ & $\avgdust$ & $\maxdust$ \\
   &  &  &  & \si{\solarmass\per\year} & \si{\solarmass\per\year} \\ \hline
  \texttt{baseline} & 0.02   & 1.20 & 1915 & \num{5.38e-10} & \num{9.06E-07} \\ \hline
  \texttt{mdot-1}   & 0.01   & 0.60 & 1915 & \num{8.79E-09} & \num{1.42E-06} \\
  \texttt{mdot-2}   & 0.04   & 2.39 & 1915 & \num{2.53E-11} & \num{5.83E-07} \\
  \texttt{mdot-3}   & 0.04   & 1.20 & 957  & \num{2.34E-08} & \num{1.17E-06} \\
  \texttt{mdot-4}   & 0.01   & 1.20 & 3830 & \num{3.81E-11} & \num{7.11E-07} \\ \hline
  \end{tabular}
  \caption{Average rate of dust production for the mass loss rate simulation set.}
  \label{tab:mdot-average-rates}
\end{table}

\subsection{Terminal velocity variation}
\label{sec:paper1vinfresults}

\begin{figure}
  \centering
  \includegraphics[width=\linewidth]{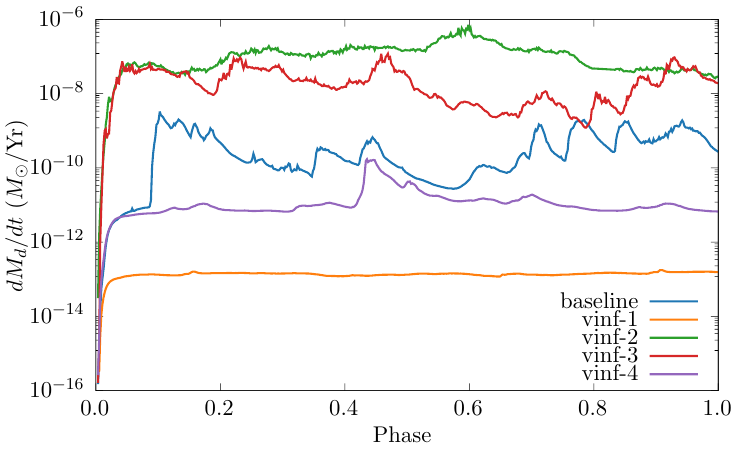}
  \caption[Comparison of the dust production rate for simulations varying the wind terminal velocity]{Comparison of the dust production rate for simulations varying the wind terminal velocity, $v^\infty$. Simulations with a strong wind velocity imbalance produce significantly more dust than their counterparts.}
  \label{fig:vinfdustproduction}
\end{figure}

\begin{figure}
  \centering
  \includegraphics[width=\linewidth]{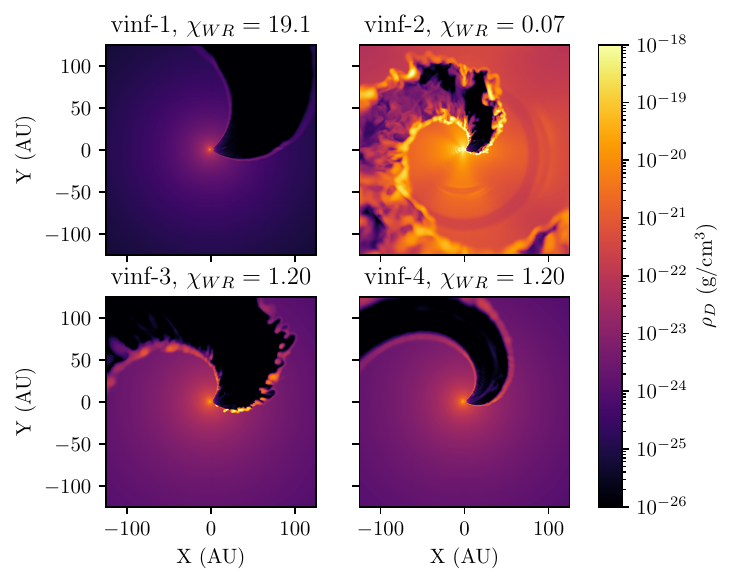}
  \caption[Dust density comparison of terminal velocity varying systems]{Comparison of the dust density in the simulations that vary $v^\infty$. Simulations with either a high OB wind velocity or low WC wind velocity produce large quantities of dust. Simulation \texttt{vinf-1}, which has a high velocity WC wind with $\chi\rms{WR} = 19.1$, does not produce any appreciable dust within the WCR. \texttt{vinf-1} and \texttt{vinf-4} have a smoother WCR with less instabilities as both winds have identical terminal speeds, resulting in no velocity shear.}
  \label{fig:vinfrhodcomp}
\end{figure}

Varying the wind terminal velocity also has an extremely strong effect on the dust formation (see Fig. \ref{fig:vinfdustproduction} and Table \ref{tab:vinf-average-rates}).
The dust production rate is exceptionally high in the case of \texttt{vinf-2}, which has an extremely slow wind velocity of \SI{500}{\kilo\metre\per\second}, closer to that of a typical LBV star than that of a WC (Table \ref{tab:vinf-average-rates}).
This very slow, dense wind experiences very strong radiative cooling in the post-shock environment ($\chi\swr = 0.07$), leading to high density pockets of cooled gas.
This can be seen in Fig. \ref{fig:vinfrhodcomp}, where \texttt{vinf-2} produces large quantities of dust near the apex of the WCR on the WR side, which is then mixed throughout the WCR.
The factor of 4 difference in the wind velocity between the WR and OB winds creates a very strong velocity shear, leading to the formation of Kelvin-Helmholtz instabilities.

It should be noted that the dust production in general increased \emph{outside} of the WCR in the case of \texttt{vinf-2} (i.e. in the unshocked WR wind).
This is largely due to the significantly higher wind density within the WC wind, and the increase in the time available for grain growth before the wind collision.
Despite this, the dust production outside of the WCR does not dominate the total dust production, most of which occurs in the WCR still. In the numerical analysis (Fig. \ref{fig:vinfdustproduction} and Table \ref{tab:vinf-average-rates}) of dust production we do not include dust produced outside of the wind collision region.
In the case of a fast WC wind (\texttt{vinf-1}) with a largely adiabatic WCR, dust production effectively ceases, with an average dust production rate of $\SI{9e-14}{\solarmass\per\year}$, two orders of magnitude less than \texttt{vinf-4}, despite the latter having a similar wind momentum ratio.

\begin{figure}
  \centering
  \includegraphics[width=\linewidth]{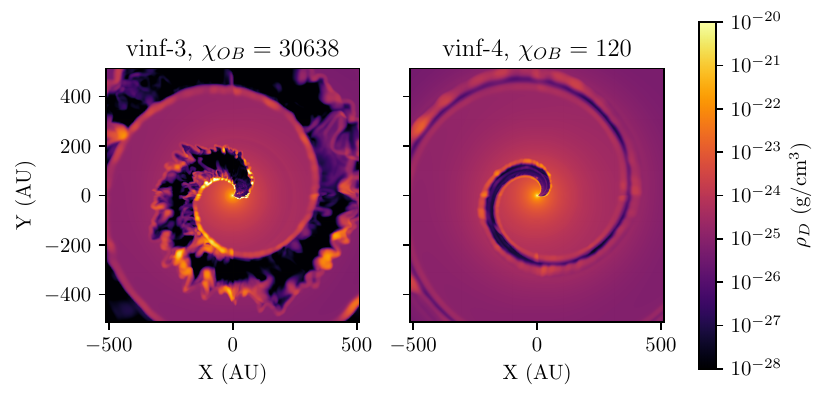}
  \caption[OB terminal velocity wind dust comparison]{Comparison of the dust density in simulations with modified OB wind terminal velocities. The simulations are fully advected with 3 orbits calculated. Dust formation and instabilities are far more pronounced in \texttt{vinf-3}, which has an OB wind velocity a factor of 4 larger than \texttt{vinf-4}.}
  \label{fig:obvinfzcomp}
\end{figure}

Simulations \texttt{vinf-3} and \texttt{vinf-4} show that when the secondary wind velocity is altered, drastic changes to the dust formation rate again occur, which is partially due to the prevalence and strength of instabilities.
A greater velocity shear along the discontinuity results in strong Kelvin-Helmholtz instabilities in \texttt{vinf-3}, whereas these are missing in \texttt{vinf-4} which has equal wind speeds.
Both \texttt{vinf-2} and \texttt{vinf-3} exhibit very strong KH instabilities, and both have a terminal velocity ratio, $v_\text{OB}^\infty / v_\text{WR}^\infty = 4$.
This augments the already present thermal instabilities due to radiative cooling, leading to a less ordered, clumpy post-shock environment.
In Fig. \ref{fig:obvinfzcomp} where \texttt{vinf-3} and \texttt{vinf-4} are directly compared, the presence of a much faster secondary wind results in a velocity shear that produces a much broader WCR, with high density pockets formed within instabilities, which appear to produce the bulk of the dust, despite both simulations having an adiabatic second wind. 
This suggests that prolific dust formation occurs in a post-shock primary wind shaped by instabilities, produced either from strong radiative cooling, or through a strong velocity shear, leading to K-H instabilities.
We note also that the dust formation rates appear to be stratified somewhat in terms of $\eta$.
Simulations where $\eta = 0.04$ produce significantly more dust than simulations with more imbalanced winds (Fig. \ref{fig:vinfdustproduction}).

\begin{figure}
  \centering
  \includegraphics[width=\linewidth]{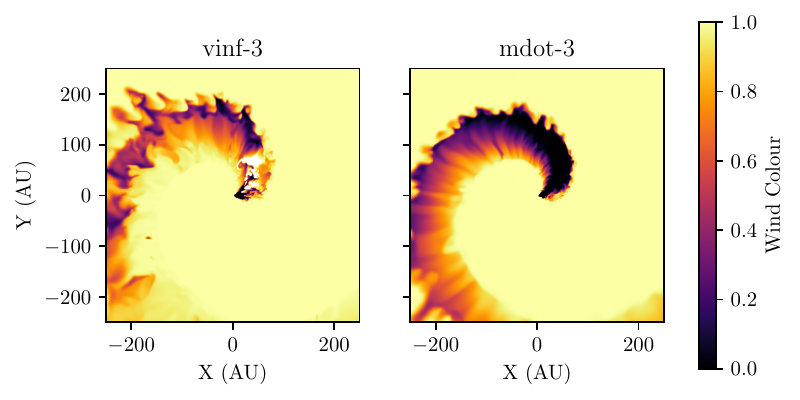}
  \caption[Wind colour comparison of $\eta = 0.04$ winds]{Comparison of the wind colour in simulations \texttt{vinf-3} and \texttt{mdot-3}. The WR wind has a colour of 1.0 while the OB wind has a colour of 0.0. Wind mixing is significantly more pronounced in \texttt{vinf-3} than in \texttt{mdot-3}. In \texttt{vinf-3} the post-shock WR wind is strongly influenced by Kelvin-Helmholtz instabilities, due to the increased wind velocity imbalance and lower degree of cooling.}
  \label{fig:eta004comparisoncolour}
\end{figure}

\begin{figure}
  \centering
  \includegraphics[width=\linewidth]{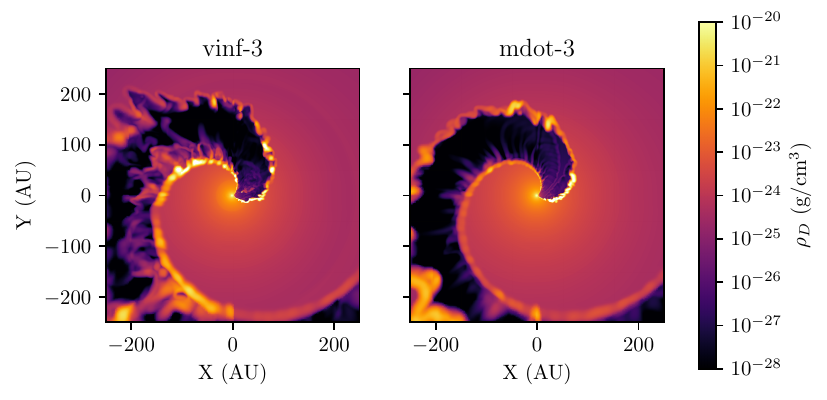}
  \caption[Comparison of dust density in simulations with strong secondary wind]{Comparison of the dust density in simulations with a strong secondary wind, models \texttt{vinf-3} and \texttt{mdot-3}. Dust in \texttt{vinf-3} is produced to a much higher degree in the trailing edge of the WCR wind, rather than on the leading edge as in \texttt{mdot-3}. The increased mixing of the winds in \texttt{vinf-3} due to Kelvin-Helmholtz instabilities has led to dust forming throughout the WCR, rather than being concentrated near the apex of the WCR.}
  \label{fig:eta004comparisonrhod}
\end{figure}

By directly comparing two prolific dust producing models with $\eta = 0.04$, models \texttt{vinf-3} and \texttt{mdot-3}, we can see that both WCRs are dominated by instabilities.
However, of the two, \texttt{vinf-3} is more thoroughly mixed (Fig. \ref{fig:eta004comparisoncolour}).
In particular, it has a much larger trailing edge that produces large quantities of dust (Fig. \ref{fig:eta004comparisonrhod}).
These simulations produce approximately the same amount of dust, with \texttt{vinf-3} also consistently producing dust in the trailing edge of the WCR.
From these results it is clear that the dust production rate is increased if there is a highly imbalanced wind velocity (with a slow WC and fast OB wind), as this leads to a post-shock environment governed by thin-shell and Kelvin-Helmholtz instabilities.

\begin{table}
  \centering
  \begin{tabular}{llllll}
  \hline
  Model & $\eta$ & $\chi_\text{WR}$ & $\chi_\text{OB}$ & $\avgdust$ & $\maxdust$ \\
   &  &  &  & \si{\solarmass\per\year} & \si{\solarmass\per\year} \\ \hline
  \texttt{baseline} & 0.02   & 1.20 & 1915  & \num{5.38e-10} & \num{9.06E-07} \\ \hline
  \texttt{vinf-1}   & 0.01   & 19.1 & 1915  & \num{8.88e-13} & \num{7.11E-07} \\
  \texttt{vinf-2}   & 0.04   & 0.07 & 1915  & \num{1.17e-7}  & \num{1.17E-06} \\
  \texttt{vinf-3}   & 0.04   & 1.20 & 30638 & \num{6.30e-11} & \num{1.17E-06} \\
  \texttt{vinf-4}   & 0.01   & 1.20 & 120   & \num{1.94e-8}  & \num{7.11E-07} \\ \hline
  \end{tabular}
  \caption{Average rates of dust production for the terminal velocity simulation set.}
  \label{tab:vinf-average-rates}
\end{table}

\subsection{Separation variation}

\begin{figure}
  \centering
  \includegraphics[width=\linewidth]{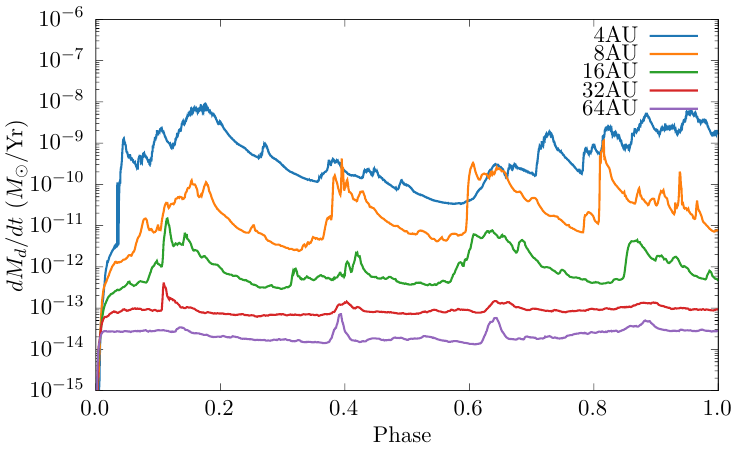}
  \caption[Dust formation rate versus binary separation distance]{A comparison of dust formation rates versus orbital phase for a set of simulations that vary the separation of the stars, $d_\text{sep}$. A clear inverse relationship between separation distance and dust production rate exists, due to the WCR behaving more adiabatically when the stars have a greater separation.}
  \label{fig:dsepdustproduction}
\end{figure}

\begin{figure*}
  \centering
  \includegraphics{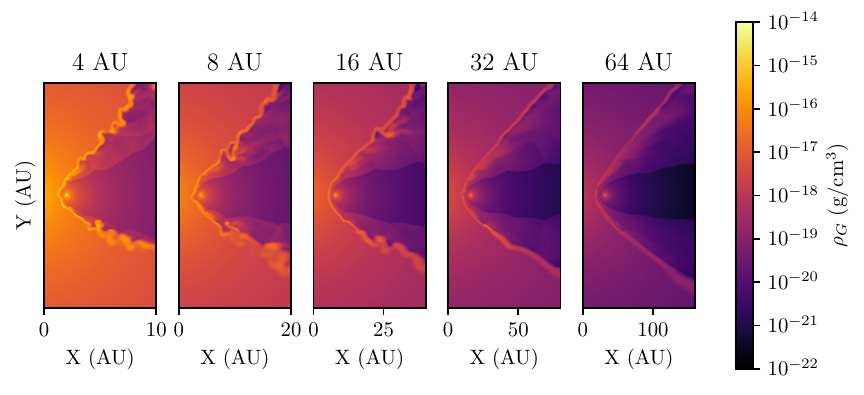}
  \caption[A comparison of the structures of simulations varying $d_\text{sep}$]{A comparison of the structures of simulations varying $d_\text{sep}$. The scale of each plot has been changed to allow for a similar feature size. Simulations with a closer stellar separation have collision regions whose structure is more strongly influenced by instabilities -- in particular by thin-shell instabilities brought on by the radiative behaviour in the WCR.}
  \label{fig:dsepinstabilities}
\end{figure*}

There is a clear correlation between the separation distance of the stars and the dust formation rate, with dust production drastically increasing as the orbital separation is decreased (Fig. \ref{fig:dsepdustproduction} and Table \ref{tab:dsep-average-rates}).
This influence on the dust formation rate is non-linear, with a doubling of the separation distance decreasing the dust production rate by approximately one order of magnitude.
At very large separations, very little dust is produced.
Clearly, dust formation is strongly influenced by the wind density at collision and the strength of the post-shock cooling. 
The variability of the dust production rate also appears to increase as the separation distance is reduced, leading to instances where a simulation may temporarily produce more dust than a simulation with a tighter orbit, such as the case with \texttt{dsep-4AU} and \texttt{dsep-8AU} at an orbital phase of $0.6 < \Phi < 0.65$.
As we have previously discussed, instabilities drive slightly intermittent, but highly efficient dust formation, which cause these fluctuations (Fig. \ref{fig:dsepinstabilities}).
Our results are consistent with observations of episodic dust forming systems, where infrared emission due to dust is maximised at or shortly after periastron passage.

\begin{table}
  \centering
  \begin{tabular}{lllll}
  \hline
  Model & $\chi\rms{WR}$ & $\chi\rms{OB}$ & $\avgdust$ & $\maxdust$ \\
   &  &  & \si{\solarmass\per\year} & \si{\solarmass\per\year} \\ \hline
  \texttt{baseline}  & 1.20 & 1915  & \num{5.38e-10} & \num{9.06E-07} \\
  \hline
  \texttt{dsep-8AU}  & 2.39 & 3830  & \num{4.39e-11} & \num{9.06E-07} \\
  \texttt{dsep-16AU} & 4.79 & 7659  & \num{1.77e-12} & \num{9.06E-07} \\
  \texttt{dsep-32AU} & 9.57 & 15319 & \num{8.83e-14} & \num{9.06E-07} \\
  \texttt{dsep-64AU} & 19.1 & 30638 & \num{2.41e-14} & \num{9.06E-07} \\ \hline
  \end{tabular}
  \caption{Average rates of dust production for the separation distance simulation set. The stellar parameters are the same as in the \texttt{baseline} model, which has a $d\rms{sep} = \SI{4.0}{\au}$.}
  \label{tab:dsep-average-rates}
\end{table}

\section{Summary \& Conclusion}
\label{sec:p1-conclusion}

Through these simulations we have how the dust yield changes for systems with differing wind and orbital parameters.
The simulations in this paper were conducted over a fairly limited parameter space of mass loss rates and wind terminal velocities.
Despite this, the dust production rate varied by up to 6 orders of magnitude.
Dust formation was found to be extremely sensitive to the wind properties of both stars, which imposes a limited range of wind parameters for dust to form efficiently.
As the stellar mass loss rates change, we find that dust production increases if either star has a greater mass loss rate; additionally systems with higher mass loss rates undergo stronger cooling.
This change could also be attributed to the increased number density of dust grains in the WC wind in the case of simulation \texttt{mdot-1}.
However, we find that a similar effect occurs when increasing the mass loss rate of the OB star, which does not inject dust into the simulation.
If the wind \emph{velocities} are more imbalanced we find that the dust production rate increases significantly as well.
This is most likely due to the presence of KH instabilities from an increased wind shear.
Dust production is also affected by the orbital properties.
As the stellar separation increases, we find a corresponding drop in the dust production rate, due to the reduction in instabilities that generate high-density clumps, as well as the reduced shock intensity as the winds become more diffuse with distance.

Our simulations explain why dust forming systems are comparatively rare, compared to the total number of systems with massive binary stars and interacting winds, and also why periodic dust forming systems have eccentric orbits.
The baseline system, which is representative of WR98a, has a significantly lower stellar mass loss rate than other well-characterised WCd systems, such as WR140 and WR104.
Future simulations will focus on these other systems
to explore how closely they match observations.
The production of dust in high density clumped regions could also explain the formation of dust in single WC stars, which can form dust in significantly lower quantities.

Furthermore, we note that the dust production rate from our \texttt{baseline} simulation is somewhat lower than the predicted dust production rate of WR98a.
A predicted value of approximately \SI{6e-7}{\solarmass\per\year} was calculated for WR98a \citep{lauRevisitingImpactDust2020}.
This discrepancy is due to a variety of factors:

\begin{itemize}
  \item Slightly different mass loss rates and wind terminal velocities of the system and the simulation. The estimated value of $v^\infty\rms{WR}$ in WR98a is 90\% of the simulation value, and as such would correspond to a significant increase in production rate based on our results.
  \item Incompleteness of the grain growth model and a lack of other growth mechanisms, such as initial growth due to impinging carbon ions.
  \item An incorrect value for the grain sticking factor, which was extremely conservative and could underestimate dust growth by a significant amount. The grain sticking factor is fairly sensitive, and would require additional research to determine an adequate value for quantitative work.
\end{itemize}

\noindent
These factors should not affect the qualitative analysis of this work, however.
A more in-depth analysis of WCd systems using an improved version of this model will allow us to make quantitative comparisons of the dust formation rate in future work.

\subsection{Wind mixing within the WCR}

While the interaction between hydrogen and dust grains is not simulated by our dust model, elements such as hydrogen are crucial for forming complex organic molecules.
As the WC wind is extremely hydrogen-poor, significant wind mixing would need to occur
\citep{herbstComplexOrganicInterstellar2009}.
Figure \ref{fig:radiative-windmixing} shows that the wind is far more effectively mixed by instabilities if it is sufficiently radiative.
An improved dust model which can calculate grain yields from chemical reactions could be used to investigate this further.
Conveniently, implementation of a chemical model into Athena++ through passive scalars is a future feature in the projects roadmap.
Additionally, a multi-fluid model could be used to model the dynamics of grains, as larger grains may not necessarily be co-moving in a turbulent wind environment.

\subsection{Summary}

Our parameter space exploration of colliding wind binary systems undergoing dust formation yields new insights into how dust forms within the WCR.
Dust production within these systems is poorly understood, and with direct observations of the WCR rendered difficult by the extreme conditions of these systems, it falls on numerical simulation to elucidate the nature of dust production in CWBs.
Our simulations reveal how sensitive to changing wind conditions this dust production is.
This parameter space exploration, whilst quite conservative, resulted in a change in dust formation rates of up to 6 orders of magnitude.
In all simulations, the bulk of dust formation was found to occur within high-density pockets formed through thin-shell or Kelvin-Helmholtz instabilities, suggesting that strong cooling and a fast secondary wind are both important factors for dust production.
For high levels of dust formation, an ideal system should have a slow, dense primary wind and a fast, dense secondary wind, with a close orbit.
This combination of properties ensures the formation of dense pockets of cool post-shock gas in which dust formation proceeds.

There is significant potential for additional research in this field.
Parameter mixing was not performed, due to the simulation time required for producing many more simulations, but performing examples on more extreme systems, such as those with a LBV primary star or a WR+WR system is a potential avenue of research.
Future work could introduce additional dust formation and destruction mechanisms, such as grain-grain collision or photodissociation.
Modelling effects such as radiative line driving or use of a multi-fluid model could also prove fruitful. 
Another interesting avenue of research is the simulation of eccentric, periodic dust forming systems; simulating either an entire or a partial orbit of a system such as WR140 would be a logical next step for this work.

\section*{Acknowledgements}

This work was undertaken on ARC4, part of the High Performance Computing facilities at the University of Leeds, UK.
We would also like to thank P. A. Crowther for his work on the Galactic Wolf-Rayet Catalogue (\url{http://pacrowther.staff.shef.ac.uk/WRcat}).
Finally, we would like to thank the academic referee for their helpful and insightful review.

\section*{Data Availability}

The data underlying this article is available in the Research Data Leeds Repository, at \url{https://doi.org/10.5518/1223}.

\bibliographystyle{mnras}
\bibliography{ref-parameter.bib} %

\bsp	%
\label{lastpage}
\end{document}